\documentclass[10pt]{article}
\usepackage[english]{babel}
\usepackage{graphics}
\usepackage{graphicx}
\usepackage{amsmath}
\usepackage{amsfonts}
\usepackage{amssymb}%
\begin{document}
\def\beq{\begin{equation}}
\def\eeq{\end{equation}}
\def\bear{\begin{eqnarray}}
\def\ear{\end{eqnarray}}
\newcommand{\Eq}[1]{Eq.\,(\ref{#1})}
\newcommand{\sect}[1]{Sec.\,#1}
\newcommand{\Ref}[1]{Ref.\,\cite{#1}}
\newcommand{\Refs}[1]{Refs.\,\cite{#1}}
\title{Primordial Black Holes}

\author{Maxim Yu. Khlopov$^{1,2,3}$\thanks{Maxim.Khlopov@roma1.infn.it}}

\date{}
\maketitle

\begin{center}
{\sl\small $^{1}$ Centre for Cosmoparticle Physics "Cosmion"\\
Moscow, 125047,
Miusskaya pl. 4 \\
$^{2}$ Moscow State Engineering Physics Institute, \\
Kashirskoe Sh., 31, Moscow 115409, Russia, and \\
$^{3}$ APC laboratory 10, rue Alice Domon et L\'eonie Duquet \\75205 Paris
Cedex 13, France}
\end{center}

\begin{abstract}
Primordial black holes (PBHs) are a profound signature of primordial
cosmological structures and provide a theoretical tool to study
nontrivial physics of the early Universe. The mechanisms of PBH
formation are discussed and observational constraints on the PBH
spectrum, or effects of PBH evaporation, are shown to restrict a
wide range of particle physics models, predicting an enhancement of
the ultraviolet part of the spectrum of density perturbations, early
dust-like stages, first order phase transitions and stages of
superheavy metastable particle dominance in the early Universe. The
mechanism of closed wall contraction can lead, in the inflationary
Universe, to a new approach to galaxy formation, involving
primordial clouds of massive BHs created around the intermediate
mass or supermassive BH and playing the role of galactic seeds.
\end{abstract}

\tableofcontents

\section{Introduction}
The convergence of the frontiers of our knowledge in micro- and
macro- worlds leads to the wrong circle of problems, illustrated by
the mystical Uhroboros (self-eating-snake). The Uhroboros puzzle may
be formulated as follows: {\it The theory of the Universe is based
on the predictions of particle theory, that need cosmology for their
test}. Cosmoparticle physics
\cite{ADS,MKH,book,book3,Khlopov:2004jb,bled} offers the way out of
this wrong circle. It studies the fundamental basis and mutual
relationship between micro-and macro-worlds in the proper
combination of physical, astrophysical and cosmological signatures.
Some aspects of this relationship, which arise in the astrophysical
problem of Primordial Black Holes (PBH) is the subject of this
review.

In particle theory Noether's theorem relates the exact symmetry to
conservation of respective charge. Extensions of the standard model
imply new symmetries and new particle states. The respective
symmetry breaking induces new fundamental physical scales in
particle theory. If the symmetry is strict, its existence implies
new conserved charge. The lightest particle, bearing this charge, is
stable. It gives rise to the deep relationship between dark matter
candidates and particle symmetry beyond the Standard model.

The mechanism of spontaneous breaking of particle symmetry also has
cosmological impact. Heating of condensed matter leads to
restoration of its symmetry. When the heated matter cools down,
phase transition to the phase of broken symmetry takes place. In the
course of the phase transitions, corresponding to given type of
symmetry breaking, topological defects can form. One can directly
observe formation of such defects in liquid crystals or in
superfluid He. In the same manner the mechanism of spontaneous
breaking of particle symmetry implies restoration of the underlying
symmetry. When temperature decreases in the course of cosmological
expansion, transitions to the phase of broken symmetry  can lead,
depending on the symmetry breaking pattern, to formation of
topological defects in very early Universe. Defects can represent
new forms of stable particles (as it is in the case of magnetic
monopoles \cite{t'Hooft,polyakov,kz,Priroda,preskill,SAOmonop}), or
extended structures, such as cosmic strings \cite{zv1,zv2} or cosmic
walls \cite{okun}.

In the old Big bang scenario  cosmological expansion and its initial
conditions were given {\it a priori} \cite{Weinberg,ZNSEU}. In the
modern cosmology expansion of Universe and its initial conditions
are related to inflation
\cite{Star80,Guthinfl,Linde:1981mu,Albrecht,Linde:1983gd},
baryosynthesis and nonbaryonic dark matter (see review in
\cite{Lindebook,Kolbbook}). Physics, underlying inflation,
baryosynthesis and dark matter, is referred to extensions of the
standard model, and variety of such extensions makes the whole
picture in general ambiguous. However, in a framework of each
particular physical realization of inflationary model with
baryosynthesis and dark matter the corresponding model dependent
cosmological scenario can be specified in all details. In such
scenario main stages of cosmological evolution, structure and
physical content of the Universe reflect structure of the underlying
physical model. The latter should include with necessity the
standard model, describing properties of baryonic matter, and its
extensions, responsible for inflation, baryosynthesis and dark
matter. In no case cosmological impact of such extensions is reduced
to reproduction of these three phenomena only. A nontrivial path of
cosmological evolution, specific for each particular realization of
inflational model with baryosynthesis and nonbaryonic dark matter,
always contains some additional model dependent cosmologically
viable predictions, which can be confronted with astrophysical data.
Here we concentrate on
 Primordial Black Holes as profound signature of such phenomena.

It was probably Pierre-Simon Laplace \cite{Laplace} in the beginning
of XIX century, who noted first that in very massive stars escape
velocity can exceed the speed of light and light can not come from
such stars. This conclusion made in the framework of Newton
mechanics and Newton corpuscular theory of light has further
transformed into the notion of "black hole" in the framework of
general relativity and electromagnetic theory. Any object of mass
$M$ can become a black hole, being put within its gravitational
radius $r_g=2 G M/c^2.$ At present time black holes (BH) can be
created only by a gravitational collapse of compact objects with
mass more than about three Solar mass \cite{1,ZNRA}. It can be a
natural end of massive stars or can result from evolution of dense
stellar clusters. However in the early Universe there were no limits
on the mass of BH. Ya.B. Zeldovich and I.D. Novikov \cite{ZN}
noticed that if cosmological expansion stops in some region, black
hole can be formed in this region within the cosmological horizon.
It corresponds to strong deviation from general expansion and
reflects strong inhomogeneity in the early Universe.  There are
several mechanisms for such strong inhomogeneity and we'll trace
their links to cosmological consequences of particle theory.


Primordial Black Holes (PBHs) are a very sensitive cosmological
probe for physics phenomena occurring in the early Universe. They
could be formed by many different mechanisms, {\it e.g.}, initial
density inhomogeneities \cite{hawking1,hawkingCarr} and non-linear
metric perturbations
\cite{Bullock:1996at,Ivanov:1997ia,Bullock:1998mi}, blue spectra of
density fluctuations
\cite{Khlopov:1984wc,polnarev,Lidsey:1995ir,Kotok:1998rp,
Dubrovich02,Sendouda:2006nu}, a softening of the equation of state
\cite{canuto,Khlopov:1984wc,polnarev}, development of gravitational
instability on early dust-like stages of dominance of supermassive
particles and scalar fields
\cite{khlopov0,polnarev0,polnarev1,khlopov1} and evolution of
gravitationally bound objects formed at these stages
\cite{Kalashnikov,Kadnikov}, collapse of cosmic strings
\cite{hawking2,Polnarev:1988dh,Hansen:2000jv,Cheng:1996du,Nagasawa:2005hv}
and necklaces \cite{Matsuda:2005ez}, a double inflation scenario
\cite{nas,Kim:1999xg,Yamaguchi:2001zh,Yamaguchi:2002sp}, first order
phase transitions \cite{hawking3,pt1Jedamzik,kkrs,kkrs1,kkrs2}, a
step in the power spectrum \cite{Sakharov0,polarski1}, etc. (see
\cite{polnarev,book,book3,Carr:2003bj,book2} for a review).

Being formed, PBHs should retain in the Universe and, if survive to
the present time, represent a specific form of dark matter
\cite{khlopov7,Ivanov:1994pa,book,book3,Blais:2002nd,Chavda:2002cj,
Afshordi:2003zb,book2,Chen:2004ft}. Effect of PBH evaporation by
S.W.Hawking \cite{hawking4} makes evaporating PBHs a source of
fluxes of products of evaporation, particularly of $\gamma$
radiation \cite{Page:1976wx}. MiniPBHs with mass below $10^{14}$~g
evaporate completely and do not survive to the present time.
However, effect of their evaporation should cause influence on
physical processes in the early Universe, thus providing a test for
their existence by methods of cosmoarcheology
\cite{Cosmoarcheology}, studying cosmological imprints of new
physics in astrophysical data. In a wide range of parameters the
predicted effect of PBHs contradicts the data and it puts
restrictions on mechanism of PBH formation and the underlying
physics of very early Universe. On the other hand, at some fixed
values of parameters, PBHs or effects of their evaporation can
provide a nontrivial solution for astrophysical problems.

Various aspects of PBH physics, mechanisms of their formation,
evolution and effects are discussed in
\cite{carr1,carrMG,LGreen,khlopov6,polnarev,Grillo:1980uj,
Chapline:1975tn,Hayward:1989jq,Yokoyama:1995ex,Kim:1996hr,
Heckler:1997jv,MacGibbon:2007yq,Page:2007yr,green,Niemeyer:1997mt,Kribs:1999bs,
Green:2004wb,Yokoyama:1998pt,Yokoyama:1998xd,Yokoyama:1999xi,
Bringmann:2001yp,Dimopoulos:2003ce,Nozari:2007kv,LythMalik,Zaballa:2006kh,
Harada:2004pf,
Custodio:2005en,Bousso:1995cc,Bousso:1996wy,Elizalde:1999dw,
Nojiri:1999vv,Bousso:1999iq,Silk:2000em,Polarski:2001jk,
Barrow:1996jk,Paul:2000jb,Paul:2001yt,Paul:2005bk,Polarski:2001yn,
Carr:1993aq,Yokoyama:1998qw,Kaloper:2004yj,Pelliccia:2007fh,
Stojkovic:2005zh,Soda,Ahn:2006uc,babichev4,babichev5,babichev6,Guariento:2007bs} 
particularly specifying PBH
formation and effects in braneworld cosmology
\cite{Guedens:2002km,Guedens:2002sd,Clancy:2003zd,
Tikhomirov:2005bt}, on inflationary preheating
\cite{Bassett:2000ha}, formation of PBHs in QCD phase transition
\cite{Jedamzik:1998hc, Widerin:1998my}, properties of superhorizon
BHs \cite{Harada:2005sc,Harada:2006gn}, role of PBHs in
baryosynthesis
\cite{Grillo:1980rt,Barrow:1990he,Turner:1979bt,Upadhyay:1999vk,
Bugaev:2001xr}, effects of PBH evaporation in the early Universe and
in modern cosmic ray, neutrino and gamma fluxes
\cite{mujana,Fegan:1978zn,Green:2001kw,Frampton:2005fk,
MacGibbon:1990zk,MacGibbon:1991tj,Halzen:1991uw,Halzen:1995hu,
Bugaev:2000bz,Bugaev:2002yt,Volkova:1994fb,Gibilisco:1996ft,
Golubkov:2000qy,He:2002vz,Gibilisco:1996dk,Custodio:2002jv,
Sendouda:2003dc,Maki:1995pa,barraupbar,Barrau:2002mc,
Wells:1998jv,Cline:1996uk,Xu:1998hn,Cline:1998fx,Sendouda:2006yc,
Barrau:1999sk,Derishev:1999xn,Tikhomirov:2004rs,Seto,barrau,
barraugamma,barrauprd}, in creation of hypothetical particles
\cite{Bell:1998jk,lemoine,green1,barrau2}, PBH clustering and
creation of supermassive BHs
\cite{Bean:2002kx,Duechting:2004dk,Chisholm:2005vm,
Dokuchaev:2004kr,Mack:2006gz,Rubin:2005pq}, effects in cosmic rays
and colliders from PBHs in low scale gravity models
\cite{barrauADDBH,barrauADDac}. Here we outline the role of PBHs as
a link in cosmoarcheoLOGICAL chain, connecting cosmological
predictions of particle theory with observational data. We discuss
the way, in which spectrum of PBHs reflects properties of superheavy
metastable particles and of phase transitions on inflationary and
post-inflationary stages. We briefly review possible cosmological
reflections of particle physics (section \ref{Cosmophenomenology}),
illustrate in section \ref{dust} some mechanisms of PBH formation on
stage of dominance of superheavy particles and fields (subsection
\ref{particles}) and from second order phase transition on
inflationary stage. Effective mechanism of BH formation during
bubble nucleation provides a sensitive tool to probe existence of
cosmological first order phase transitions by PBHs (section
\ref{phasetransitions}). Existence of stable remnants of PBH
evaporation can strongly increase the sensitivity of such probe and
we demonstrate this possibility in section \ref{gravitino} on an
example of gravitino production in PBH evaporation. Being formed
within cosmological horizon, PBHs seem to have masses much less than
the mass of stars, constrained by small size of horizon in very
early Universe. However, if phase transition takes place on
inflationary stage, closed walls of practically any size can be
formed (subsection \ref{walls}) and their successive collapse can
give rise to clouds of massive black holes, which can play the role
of seeds for galaxies (section \ref{MBHwalls}). The impact of
constraints and cosmological scenarios, involving primordial black
holes, is briefly discussed in section \ref{Discussion}.

\section{PBHs as cosmological reflection of new physics}\label{Cosmophenomenology}
The simplest primordial form of new physics is a gas of new stable
massive particles, originated from early Universe. For particles
with mass $m$, at high temperature $T>m$ the equilibrium condition,
$n \cdot \sigma v \cdot t > 1$ is valid, if their annihilation cross
section $\sigma > 1/(m m_{pl})$ is sufficiently large to establish
equilibrium. At $T<m$ such particles go out of equilibrium and their
relative concentration freezes out. Weakly interacting species
decouple from plasma and radiation at $T>m$, when $n \cdot \sigma v
\cdot t \sim 1$, i.e. at $T_{dec} \sim (\sigma m_{pl})^{-1}$. This
is the main idea of calculation of primordial abundance for
WIMP-like dark matter candidates (see e.g.
\cite{book,book3,Cosmoarcheology} for details). The maximal
temperature, which is reached in inflationary Universe, is the
reheating temperature, $T_{r}$, after inflation. So, very weakly
interacting particles with annihilation cross section $\sigma <
1/(T_{r} m_{pl})$, as well as very heavy particles with mass $m \gg
T_{r}$ can not be in thermal equilibrium, and the detailed mechanism
of their production should be considered to calculate their
primordial abundance.

Decaying particles with lifetime $\tau$, exceeding the age of the
Universe, $t_{U}$, $\tau > t_{U}$, can be treated as stable. By
definition, primordial stable particles survive to the present time
and should be present in the modern Universe. The net effect of
their existence is given by their contribution into the total
cosmological density. They can dominate in the total density being
the dominant form of cosmological dark matter, or they can represent
its subdominant fraction. In the latter case more detailed analysis
of their distribution in space, of their condensation in galaxies,
of their capture by stars, Sun and Earth, as well as effects of
their interaction with matter and of their annihilation provides
more sensitive probes for their existence. In particular,
hypothetical stable neutrinos of 4th generation with mass about 50
GeV are predicted to form the subdominant form of modern dark
matter, contributing less than 0,1 \% to the total density
\cite{ZKKC,DKKM}. However, direct experimental search for cosmic
fluxes of weakly interacting massive particles (WIMPs) may be
sensitive to existence of such component (see
\cite{DAMA,DAMA-review,CDMS,CDMS2} and references therein). It was
shown in \cite{Fargion99,Grossi,Belotsky,Belotsky2} that
annihilation of 4th neutrinos and their antineutrinos in the Galaxy
can explain the galactic gamma-background, measured by EGRET in the
range above 1 GeV, and that it can give some clue to explanation of
cosmic positron anomaly, claimed to be found by HEAT. 4th neutrino
annihilation inside the Earth should lead to the flux of underground
monochromatic neutrinos of known types, which can be traced in the
analysis of the already existing and future data of underground
neutrino detectors \cite{Belotsky,BKS1,BKS2,BKS3}.

New particles with electric charge and/or strong interaction can
form anomalous atoms and contain in the ordinary matter as anomalous
isotopes. For example, if the lightest quark of 4th generation is
stable, it can form stable charged hadrons, serving as nuclei of
anomalous atoms of e.g. anomalous helium
\cite{BKS,BKSR,BKSR1,BKSR2,BKSR3,BKSR4}.

Primordial unstable particles with lifetime, less than the age of
the Universe, $\tau < t_{U}$, can not survive to the present time.
But, if their lifetime is sufficiently large to satisfy the
condition $\tau \gg (m_{pl}/m) \cdot (1/m)$, their existence in
early Universe can lead to direct or indirect traces. Cosmological
flux of decay products contributing into the cosmic and gamma ray
backgrounds represents the direct trace of unstable particles. If
the decay products do not survive to the present time their
interaction with matter and radiation can cause indirect trace in
the light element abundance or in the fluctuations of thermal
radiation.

If particle lifetime is much less than $1$ s multi-step indirect
traces are possible, provided that particles dominate in the
Universe before their decay. On dust-like stage of their dominance
black hole formation takes place, and spectrum of such primordial
black holes traces particle properties (mass, frozen concentration,
lifetime) \cite{polnarev}. Particle decay in the end of dust like
stage influences the baryon asymmetry of the Universe. In any way
cosmophenomenoLOGICAL chains link the predicted properties of even
unstable new particles to the effects accessible in astronomical
observations. Such effects may be important in analysis of the
observational data.

Parameters of new stable and metastable particles are also
determined by a pattern of particle symmetry breaking. This pattern
is reflected in a succession of phase transitions in the early
Universe. First order phase transitions proceed through bubble
nucleation, which can result in black hole formation (see e.g.
\cite{kkrs} and \cite{book2} for review and references). Phase
transitions of the second order can lead to formation of topological
defects, such as walls, string or monopoles. The observational data
put severe constraints on magnetic monopole \cite{kz} and cosmic
wall production \cite{okun}, as well as on the parameters of cosmic
strings \cite{zv1,zv2}. Structure of cosmological defects can be
changed in succession of phase transitions. More complicated forms
like walls-surrounded-by-strings can appear. Such structures can be
unstable, but their existence can leave a trace in nonhomogeneous
distribution of dark matter and give rise to large scale structures
of nonhomogeneous dark matter like {\it archioles}
\cite{Sakharov2,kss,kss2}. Primordial Black Holes represent a
profound signature of such structures.
\section{PBHs from early dust-like stages}\label{dust}
A possibility to form a black hole is highly improbable in
homogeneous expanding Universe, since it implies metric fluctuations
of order 1. For metric fluctuations distributed according to
Gaussian law with dispersion \begin{equation}
\label{DispBH}\left\langle \delta^2 \right\rangle \ll
1\end{equation} a probability for fluctuation of order 1 is
determined by exponentially small tail of high amplitude part of
this distribution. This probability can be even more suppressed in a
case of non-Gaussian flutuations \cite{Bullock:1996at}.

In the Universe with equation of state \begin{equation}
\label{EqState}p=\gamma \epsilon,\end{equation} with numerical
factor $\gamma$ being in the range \begin{equation}
\label{FacState}0 \le \gamma \le 1\end{equation} a probability to
form black hole from fluctuation within cosmological horizon is
given by (see e.g. \cite{book,book3} for review and references)
\begin{equation}
\label{ProbBH}W_{PBH} \propto \exp \left(-\frac{\gamma^2}{2
\left\langle \delta^2 \right\rangle}\right).
\end{equation}
It provides exponential sensitivity of PBH spectrum to softening of
equation of state in early Universe ($\gamma \rightarrow 0$) or to
increase of ultraviolet part of spectrum of density fluctuations
($\left\langle \delta^2 \right\rangle \rightarrow 1$). These
phenomena can appear as cosmological consequence of particle theory.
\subsection{Dominance of superheavy particles in early Universe}\label{particles}
Superheavy particles can not be studied at accelerators directly. If
they are stable, their existence can be probed by cosmological
tests, but there is no direct link between astrophysical data and
existence of superheavy metastable particles with lifetime $\tau \ll
1s$. It was first noticed in \cite{khlopov0} that dominance of such
particles in the Universe before their decay at $t \le \tau$ can
result in formation of PBHs, retaining in Universe after the
particles decay and keeping some information on particle properties
in their spectrum. It provided though indirect but still a
possibility to probe existence of such particles in astrophysical
observations. Even the absence of observational evidences for PBHs
is important. It puts restrictions on allowed properties of
superheavy metastable particles, which might form such PBHs on a
stage of particle dominance, and thus constrains parameters of
models, predicting these particles.

After reheating, at \begin{equation} \label{Eareq}T <
T_0=rm\end{equation} particles with mass $m$ and relative abundance
$r=n/n_r$ (where $n$ is frozen out concentration of particles and
$n_r$ is concentration of relativistic species) must dominate in the
Universe before their decay. Dominance of these nonrelativistic
particles at $t>t_0$, where
\begin{equation}
\label{EarMD}t_0=\frac{m_{pl}}{T_0^2},\end{equation} corresponds to
dust like stage with equation of state $p=0,$ at which particle
density fluctuations grow as\begin{equation}
\label{dens}\delta(t)=\frac{\delta \rho}{\rho} \propto t^{2/3}
\end{equation}
 and development of gravitational instability results in formation
of gravitationally bound systems, which decouple at \begin{equation}
\label{decoup}t \sim t_f \approx t_i
\delta(t_i)^{-3/2}\end{equation} from general cosmological
expansion, when $\delta(t_f)\sim 1$ for fluctuations, entering
horizon at $t=t_i>t_0$ with amplitude $\delta(t_i)$.

Formation of these systems can result in black hole formation either
immediately after the system decouples from expansion or in result
of evolution of initially formed nonrelativistic gravitationally
bound system.

If density fluctuation is especially homogeneous and isotropic, it
directly collapses to BH as soon as the amplitude of fluctuation
grows to 1 and the system decouples from expansion. A probability
for direct BH formation in collapse of such homogeneous and
isotropic configurations gives minimal estimation of BH formation on
dust-like stage.

This probability was calculated in \cite{khlopov0} with the use of
the following arguments. In the period $t \sim t_f$, when
fluctuation decouples from expansion, its configuration is defined
by averaged density $\rho_1$, size $r_1$, deviation from sphericity
$s$ and by inhomogeneity $u$ of internal density distribution within
the fluctuation.  Having decoupled from expansion, the configuration
contracts and the minimal size to which it can contract is
\begin{equation} \label{sphcontr}r_{min} \sim s r_1,\end{equation}
being determined by a deviation from sphericity
\begin{equation} \label{spheric}s=\max\{\left\vert\gamma_1-\gamma_2\right\vert,\left\vert\gamma_1-
\gamma_3\right\vert,\left\vert\gamma_3-\gamma_2\right\vert\},\end{equation}
where $\gamma_1$, $\gamma_2$ and $\gamma_3$ define a deformation of
configuration along its three main orthogonal axes. It was first
noticed in \cite{khlopov0} that to form a black hole in result of
such contraction it is sufficient that configuration returns to the
size \begin{equation} \label{rminBH}r_{min} \sim r_g \sim t_i \sim
\delta(t_i) r_1,\end{equation} which had the initial fluctuation
$\delta(t_i)$, when it entered horizon at cosmological time $t_i$.
If
\begin{equation} \label{spher}s \le \delta(t_i),\end{equation}
configuration is sufficiently isotropic to concentrate its mass in
the course of collapse within its gravitational radius, but such
concentration also implies sufficient homogeneity of configuration.
Density gradients can result in gradients of pressure, which can
prevent collapse to BH. This effect does not take place for
contracting collisionless gas of weakly interacting massive
particles, but due to inhomogeneity of collapse the particles, which
have already passed the caustics can free stream beyond the
gravitational radius, before the whole mass is concentrated within
it. Collapse of nearly spherically symmetric dust configuration is
described by Tolmen solution. It's analysis
\cite{polnarev0,polnarev1,KP,polnarev} has provided a constraint on
the inhomogeneity $u=\delta \rho_1/\rho_1$ within the configuration.
It was shown that both for collisionless and interacting particles
the condition \begin{equation}
\label{inhom}u<\delta(t_i)^{3/2}\end{equation} is sufficient for
configuration to contract within its gravitational radius.

A probability for direct BH formation is then determined by a
product of probability for sufficient initial sphericity $W_s$ and
homogeneity $W_u$ of configuration, which is determined by the phase
space for such configurations. In a calculation of $W_s$ one should
take into account that the condition (\ref{spher}) implies 5
conditions for independent components of tensor of deformation
before its diagonalization (2 conditions for three diagonal
components to be close to each other and 3 conditions for
nondiagonal components to be small). Therefore, the probability of
sufficient sphericity is given by
\cite{khlopov0,polnarev0,polnarev1,KP,polnarev} \begin{equation}
\label{Wspher}W_s \sim \delta(t_i)^{5}\end{equation} and together
with the probability for sufficient homogeneity \begin{equation}
\label{Winhom}W_u \sim \delta(t_i)^{3/2}\end{equation} results in
the strong power-law suppression of probability for direct BH
formation \beq \label{WPBH} W_{PBH} = W_s \cdot W_u \sim
\delta(t_i)^{13/2}.\eeq Though this calculation was originally done
in \cite{khlopov0,polnarev0,polnarev1,KP,polnarev} for Gaussian
distribution of fluctuations, it does not imply specific form of
high amplitude tail of this distribution and thus should not change
strongly in a case of non-Gaussian fluctuations
\cite{Bullock:1996at}.

The mechanism
\cite{khlopov0,polnarev0,polnarev1,KP,polnarev,book,book3} is
effective for formation of PBHs with mass in an interval \beq
\label{Mint}M_0 \le M \le M_{bhmax}.\eeq The minimal mass
corresponds to the mass within cosmological horizon in the period $t
\sim t_0,$ when particles start to dominate in the Universe and it
is equal to
\cite{khlopov0,polnarev0,polnarev1,KP,polnarev,book,book3}\beq
\label{MBHmin} M_{0} = \frac{4 \pi}{3} \rho t^3_0 \approx
m_{pl}(\frac{m_{pl}}{r m})^2.\eeq
 The maximal mass is indirectly determined by the condition
 \beq
\label{Mconmax}\tau = t(M_{bhmax}) \delta(M_{bhmax})^{-3/2}\eeq that
fluctuation
 in the considered scale $M_{bhmax}$, entering the horizon at $t(M_{bhmax})$
 with an amplitude $\delta(M_{bhmax})$ can manage to grow up to nonlinear stage,
 decouple and collapse before particles decay at $t=\tau.$
 For scale invariant spectrum $\delta(M)=\delta_0$ the maximal mass
 is given by
\cite{book2}\beq \label{MBHmax} M_{bhmax} = m_{pl}
\frac{\tau}{t_{Pl}} \delta_0^{3/2} =m_{pl}^2 \tau
\delta_0^{3/2}.\eeq The probability, given by Eq.(\ref{WPBH}), is
also appropriate for formation of PBHs on dust-like preheating stage
after inflation \cite{khlopov1,book,book3}. The simplest example of
such stage can be given with the use of a model of homogeneous
massive scalar field \cite{book,book3}. Slow rolling of the field in
the period $t \ll 1/m$ (where $m$ is the mass of field) provides
chaotic inflation scenario, while at $t
> 1/m$ the field oscillates with period $1/m$. Coherent oscillations
of the field correspond to an averaged over period of oscillations
dust-like equation of state $p=0,$ at which gravitational
instability can develop. The minimal mass in this case corresponds
to the Jeans mass of scalar field, while the maximal mass is also
determined by a condition that fluctuation grows and collapses
before the scalar field decays and reheats the Universe.

The probability $W_{PBH}(M)$ determines the fraction of total
density \beq \label{beta}\beta(M)=\frac{\rho_{PBH}(M)}{\rho_{tot}}
\approx W_{PBH}(M),\eeq corresponding to PBHs with mass $M$. For
$\delta(M) \ll 1$ this fraction, given by Eq.(\ref{WPBH}), is small.
It means that the bulk of particles do not collapse directly in
black holes, but form gravitationally bound systems. Evolution of
these systems can give much larger amount of PBHs, but it strongly
depends on particle properties.

Superweakly interacting particles form gravitationally bound systems
of collisionless gas, which remind modern galaxies with
collisionless gas of stars. Such system can finally collapse to
black hole, but energy dissipation in it and consequently its
evolution is a relatively slow process \cite{ZPod,book,book3}. The
evolution of these systems is dominantly determined by evaporation
of particles, which gain velocities, exceeding the parabolic
velocity of system. In the case of binary collisions the evolution
timescale can be roughly estimated \cite{ZPod,book,book3} as \beq
\label{tevbin} t_{ev} = \frac{N}{\ln{N}} t_{ff}\eeq for
gravitationally bound system of $N$ particles, where the free fall
time $t_{ff}$ for system with density $\rho$ is $t_{ff} \approx (4
\pi G \rho)^{-1/2}.$ This time scale can be shorter due to
collective effects in collisionless gas \cite{GurSav} and be at
large $N$ of the order of \beq \label{tevcol} t_{ev} \sim N^{2/3}
t_{ff}.\eeq However, since the free fall time scale for
gravitationally bound systems of collisionless gas is of the order
of cosmological time $t_f$ for the period, when these systems are
formed, even in the latter case the particles should be very long
living $\tau \gg t_f$ to form black holes in such slow evolutional
process.

The evolutional time scale is much smaller for gravitationally bound
systems of superheavy particles, interacting with light relativistic
particles and radiation. Such systems have analogy with stars, in
which evolution time scale is defined by energy loss by radiation.
An example of such particles give superheavy color octet fermions of
asymptotically free SU(5) model \cite{Kalashnikov} or magnetic
monopoles of GUT models. Having decoupled from expansion, frozen out
particles and antiparticles can annihilate in gravitationally bound
systems, but detailed numerical simulation  \cite{Kadnikov} has
shown that annihilation can not prevent collapse of the most of mass
and the timescale of collapse does not exceed the cosmological time
of the period, when the systems are formed.

\subsection{Spikes from phase transitions on inflationary stage}
Scale non-invariant spectrum of fluctuations, in which amplitude of
small scale fluctuations is enhanced, can be another factor,
increasing the probability of PBH formation. The simplest functional
form of such spectrum is represented by a blue spectrum with a power
law dispersion \beq\left\langle \delta^2(M) \right\rangle \propto
M^{-k},\eeq with amplitude of fluctuations growing at $k>0$ to small
$M$. The realistic account for existence of other scalar fields
together with inflaton in the period of inflation can give rise to
spectra with distinguished scales, determined by parameters of
considered fields and their interaction.

In chaotic inflation scenario interaction of a Higgs field $\phi$
with inflaton $\eta$ can give rise to phase transitions on
inflationary stage, if this interaction induces positive mass term
$+\frac {\nu^2}{2} \eta^2 \phi^2$. When in the course of slow
rolling the amplitude of inflaton decreases below a certain critical
value $\eta_c = m_{\phi}/\nu$ the mass term in Higgs potential \beq
\label{Higgs} V(\phi, \eta)=-
\frac{m^2_{\phi}}{2}\phi^2+\frac{\lambda_{\phi}}{4}\phi^4 +\frac
{\nu^2}{2}\eta^2 \phi^2 \eeq changes sign and phase transition takes
place. Such phase transitions on inflationary stage lead to the
appearance of a characteristic spikes in the spectrum of initial
density perturbations. These spike--like perturbations re-enter 
the horizon during the radiation or dust like era
and could in principle collapse to form primordial black holes. The
possibility of such spikes in chaotic inflation scenario was first
pointed out in \cite{KofLin} and realized  in \cite{Sakharov0} as a
mechanism of of PBH formation for the model of horizontal
unification \cite{Berezhiani1,Berezhiani2,Berezhiani3,Sakharov1}.

For vacuum expectation value of a Higgs field \beq\left\langle \phi
\right\rangle = \frac{m}{\lambda} = v\eeq and $\lambda \sim 10^{-3}$
the amplitude $\delta$ of spike in spectrum of density fluctuations,
generated in phase transition on inflationary stage is given by
\cite{Sakharov0} \beq \label{dspike}\delta \approx \frac{4}{9s}\eeq
with \beq \label{spike} s=\sqrt{ \frac{4}{9}+\kappa
10^5\left(\frac{v}{m_{pl}}\right)^2}-\frac{3}{2}, \eeq where $\kappa
\sim 1.$

If phase transition takes place at $e$--folding $N$ before the end
of inflation and the spike re-enters horizon on radiation dominance
(RD) stage, it forms Black hole of mass \beq \label{Mrd} M \approx
\frac{m^2_{Pl}}{H_0} \exp\{2 N\}, \eeq where $H_0$ is the Hubble
constant in the period of inflation.

If the spike re-enters horizon on matter dominance (MD) stage it
should form black holes of mass \beq \label{Mmd} M \approx
\frac{m^2_{Pl}}{H_0} \exp\{3 N\}. \eeq

\section{First order
phase transitions as a source of black holes in the early
Universe}\label{phasetransitions} First order phase transition go
through bubble nucleation. Remind the common example of boiling
water. The simplest way to describe first order phase transitions
with bubble creation in early Universe is based on a scalar field
theory with two non degenerated vacuum states. Being stable at a
classical level, the false vacuum state decays due to quantum
effects, leading to a nucleation of bubbles of true vacuum and their
subsequent expansion \cite {6}. The potential energy of the false
vacuum is converted into a kinetic energy of bubble walls thus
making them highly relativistic in a short time. The bubble expands
till it collides with another one. As it was shown in
\cite{hawking3,5} a black hole may be created in a collision of
several bubbles. The probability for collision of two bubbles is
much higher. The opinion of the BH absence in such processes was
based on strict conservation of the original O(2,1) symmetry. As it
was shown in \cite{kkrs,kkrs1,kkrs2} there are ways to break it.
Firstly, radiation of scalar waves indicates the entropy increasing
and hence the permanent breaking of the symmetry during the bubble
collision. Secondly, the vacuum decay due to thermal fluctuation
does not possess this symmetry from the beginning. The
investigations \cite{kkrs,kkrs1,kkrs2} have shown that BH can be
created as well with a probability of order unity in collisions of
only two bubbles. It initiates an enormous production of BH that
leads to essential cosmological consequences discussed below.

In subsection \ref{field} the evolution of the field configuration
in the collisions of bubbles is discussed. The BH mass distribution
is obtained in subsection \ref{Collapse}. In subsection \ref{pt1}
cosmological consequences of BH production in bubble collisions at
the end of inflation are considered.

\subsection{Evolution of field configuration in collisions of vacuum
bubbles}\label{field}

Consider a theory where a probability of false vacuum decay equals
$\Gamma $ and difference of energy density between the false and
true vacuum outside equals $\rho_v$. Initially bubbles are produced
at rest however walls of the bubbles quickly increase their velocity
up to the speed of light $v=c=1$ because a conversion of the false
vacuum energy into its kinetic ones is energetically favorable.

Let us discuss dynamics of collision of two true vacuum bubbles that
have been nucleated in points $({\bf r}_1,t_1),({\bf r}_2,t_2)$ and
which are expanding into false vacuum. Following papers
\cite{hawking3,7} let us assume for simplicity that the horizon size
is much greater than the distance between the bubbles. Just after
collision mutual penetration of the walls up to the distance
comparable with its width is accompanied by a significant potential
energy increase \cite{8}. Then the walls reflect and accelerate
backwards. The space between them is filled by the field in the
false vacuum state converting the kinetic energy of the wall back to
the energy of the false vacuum state and slowdown the velocity of
the walls. Meanwhile the outer area of the false vacuum is absorbed
by the outer wall, which expands and accelerates outwards.
Evidently, there is an instant when the central region of the false
vacuum is separated. Let us note this false vacuum bag (FVB) does
not possess spherical symmetry at the moment of its separation from
outer walls but wall tension restores the symmetry during the first
oscillation of FVB. As it was shown in \cite{7}, the further
evolution of FVB consists of several stages:

1) FVB grows up to the definite size $D_M$ until the kinetic energy
of its wall becomes zero;

2) After this moment the false vacuum bag begins to shrink up to a
minimal size $D^{*}$;

3) Secondary oscillation of the false vacuum bag occurs.

The process of periodical expansions and contractions leads to
energy losses of FVB in the form of quanta of scalar field. It has
been shown in the \cite {7,9} that only several oscillations take
place. On the other hand, important note is that the secondary
oscillations might occur only if the minimal size of the FVB would
be larger than its gravitational
radius, $%
D^{*}>r_g$. Then oscillating solutions of "quasilumps" can be
realized \cite {oscilon}. The opposite case ($D^{*}<r_g$ ) leads to
a BH creation with the mass about the mass of the FVB. As it was
shown in \cite{kkrs,kkrs1,kkrs2} the probability of BH formation is
almost unity in a wide range of parameters of theories with first
order phase transitions.

\subsection{Gravitational collapse of FVB and BH creation}\label{Collapse}

Consider following \cite{kkrs,kkrs1,kkrs2,book2,book3} in more
details the conditions of converting FVB into BH. The mass $M$ of
FVB can be calculated in a framework of a specific theory and can be
estimated in a coordinate system $K^{\prime }$ where the colliding
bubbles are nucleated simultaneously. The radius of each bubble
$b^{\prime}$ in this system equals to half of their initial
coordinate distance at first moment of collision. Apparently the
maximum size $D_M$ of the FVB is of the same order as the size of
the bubble, since this is the only parameter of necessary dimension
on such a scale: $D_M=2b^{\prime }C$. The parameter $C\simeq 1$ is
obtained by numerical calculations in the framework of each theory,
but its exact numerical value does not affect significantly
conclusions.

One can find the mass of FVB that arises at the collision of two
bubbles of radius:

\begin{equation}
\label{one}M=\frac{4\pi }3\left( Cb^{\prime }\right) ^3\rho_v
\end{equation}
This mass is contained in the shrinking area of false vacuum.
Suppose for estimations that the minimal size of FVB is of order of
wall width $\Delta $. The BH is created if minimal size of FVB is
smaller than its gravitational radius. It means that at least at the
condition

\begin{equation}
\label{two}\Delta <r_g=2GM
\end{equation}
the FVB can be converted into BH (where G is the gravitational
constant).

As an example consider a simple model with Lagrangian

\begin{equation}
\label{three}L=\frac 12\left( \partial _\mu \Phi \right) ^2-\frac
\lambda 8\left( \Phi ^2-\Phi _0^2\right) ^2-\epsilon \Phi _0^3\left(
\Phi +\Phi _0\right) .
\end{equation}
In the thin wall approximation the width of the bubble wall can be
expressed as $\Delta =2\left( \sqrt{\lambda }\Phi _0\right) ^{-1}$.
Using (\ref{two}) one can easily derive that at least FVB with mass

\begin{equation}
\label{four}M>\frac 1{\sqrt{\lambda }\Phi _0G}
\end{equation}
should be converted into BH of mass M. The last condition is valid
only in case when FVB is completely contained within the
cosmological horizon,
namely $%
M_H>1/\sqrt{\lambda }\Phi _0G$ where the mass of the cosmological
horizon at the moment of phase transition is given by $M_H\cong
m_{pl}^3/\Phi_{0}^2$. Thus for the potential (\ref{three}) at the
condition $\lambda >(\Phi_0/m_{pl})^2$ a BH is formed. This condition
is valid for any realistic set of parameters of theory.

The mass and velocity distribution of FVBs, supposing its mass is
large enough to satisfy the inequality (\ref{two}), has been found
in \cite{kkrs,kkrs1,kkrs2}. This distribution can be written in the
terms of dimensionless mass $\mu \equiv \left( \frac \pi 3\Gamma
\right) ^{1/4}\left( \frac M{C\rho _v}\right) ^{1/3}$:

\begin{equation}
\label{12}
\begin{array}{c}
\frac{dP}{\Gamma ^{-3/4}Vdvd\mu }=64\pi \left( \frac \pi 3\right)
^{1/4}\mu ^3e^{\mu ^4}\gamma ^3J(\mu ,v), \\ J(\mu ,v)=\int_{\tau
_{}}^\infty d\tau e^{-\tau ^4},\tau _{-}=\mu \left[ 1+\gamma
^2\left( 1+v\right) \right] .
\end{array}
\end{equation}

The numerical integration of (\ref{12}) revealed that the
distribution is rather narrow. For example the number of BH with
mass 30 times greater than the average one is suppressed by factor
$10^5$. Average value of the non dimensional mass is equal to
$\mu=0.32$. It allows to relate the average mass of BH and volume
containing the BH at the moment of the phase transition:

\begin{equation}
\label{MV}\left\langle M_{BH}\right\rangle =\frac C4\mu ^3\rho
_v\left\langle V_{BH}\right\rangle \simeq 0.012\rho _v\left\langle
V_{BH}\right\rangle .
\end{equation}

\subsection{First order phase transitions in the early Universe}\label{pt1}

Inflation models ended by a first order phase transition hold a
dignified position in the modern cosmology of early Universe (see
for example \cite{10,101,102,103,104,11,111}). The interest to these
models is due to, that such models are able to generate the observed
large-scale voids as remnants of the primordial bubbles for which
the characteristic wavelengths are several tens of Mpc.
\cite{11,111}. A detailed analysis of a first order phase transition
in the context of extended inflation can be found in \cite{12}.
Hereafter we will be interested only in a final stage of inflation
when the phase transition is completed. Remind that a first order
phase transition is considered as completed immediately after
establishing of true vacuum percolation regime. Such regime is
established approximately when at least one bubble per unit Hubble
volume is nucleated. Accurate computation \cite{12} shows that first
order phase transition is successful if the following condition is
valid:
\begin{equation}
\label{14}Q\equiv \frac{4\pi }9\left( \frac \Gamma {H^4}\right)
_{t_{end}}=1.
\end{equation}
Here $\Gamma$ is the bubble nucleation rate. In the framework of
first order inflation models the filling of all space by true vacuum
takes place due to bubble collisions, nucleated at the final moment
of exponential expansion. The collisions between such bubbles occur
when they have comoving spatial dimension less or equal to the
effective Hubble horizon $H_{end}^{-1}$ at the transition epoch. If
we take $H_0=100hKm/\sec /Mpc$ in $\Omega =1$
Universe the comoving size of these bubbles is approximately $%
10^{-21}h^{-1}Mpc$. In the standard approach it believes that such
bubbles are rapidly thermalized without leaving a trace in the
distribution of matter and radiation. However, in the previous
subsection it has been shown that for any realistic parameters of
theory, the collision between only two bubble leads to BH creation
with the probability closely to 100\% . The mass of this BH is given
by (see (\ref{MV}))
\begin{equation}
\label{15}M_{BH}=\gamma _1M_{bub}
\end{equation}
where $\gamma _1\simeq 10^{-2}$ and $M_{bub}$ is the mass that could
be contained in the bubble volume at the epoch of collision in the
condition of a full thermalization of bubbles. The discovered
mechanism leads to a new direct possibility of PBH creation at the
epoch of reheating in first order inflation models. In standard
picture PBHs are formed in the early Universe if density
perturbations are sufficiently large, and the probability of PBHs
formation from small post- inflation initial perturbations is
suppressed (see Section \ref{dust}). Completely different situation
takes place at final epoch of first order inflation stage; namely
collision between bubbles of Hubble size in percolation regime leads
to copious PBH formation with masses

\begin{equation}
\label{16}M_0=\gamma _1M_{end}^{hor}= \frac{\gamma
_1}2\frac{m_{pl}^2}{H_{end}},
\end{equation}
where $M_{end}^{hor}$ is the mass of Hubble horizon at the end of
inflation. According to (\ref{MV}) the initial mass fraction of this
PBHs is given by $\beta _0\approx\gamma _1/e\approx 6\cdot 10^{-
3}$. For example, for typical value of $H_{end}\approx 4\cdot
10^{-6}m_{pl}$ the initial mass fraction $\beta $ is contained in
PBHs with mass $M_0\approx 1g$.

In general the Hawking evaporation of mini BHs could give rise to a
variety possible end states. It is generally assumed, that
evaporation proceeds until the PBH vanishes completely \cite{21},
but there are various arguments against this proposal (see e.g.
\cite{22,carr1,222,223}). If one supposes that BH evaporation leaves
a stable relic, then it is naturally to assume that it has a mass of
order $m_{rel}=km_{pl}$, where $k\simeq 1\div 10^2$. We can
investigate the consequences of PBH forming at the percolation epoch
after first order inflation, supposing that the stable relic is a
result of its evaporation. As it follows from the above
consideration the PBHs are preferentially formed with a typical mass
$M_0$ at a single time $t_1$. Hence the total density $\rho$ at this
time is
\begin{equation}
\label{totdens} \rho (t_1)=\rho_{\gamma}(t_1)+\rho_{PBH}(t_1)=
\frac{3(1-\beta_0)}{32\pi t_1^2}m_{pl}^2+ \frac{3\beta_0}{32\pi
t_1^2}m_{pl}^2,
\end{equation}
where $\beta_0$ denotes the fraction of the total density,
corresponding to PBHs in the period of their formation $t_1$. The
evaporation time scale can be written in the following form
\begin{equation}
\label{evop} \tau_{BH}=\frac{M_0^3}{g_*m_{pl}^4}
\end{equation}
where $g_*$ is the number of effective massless degrees of freedom.

Let us derive the density of PBH relics. There are two distinct
possibilities to consider.

The Universe is still radiation dominated (RD) at $\tau_{BH}$. This
situation will be hold if the following condition is valid
$\rho_{BH}(\tau_{BH})<\rho_{\gamma}(\tau_{BH})$. It is possible to
rewrite this condition in terms of Hubble constant at the end of
inflation
\begin{equation}
\label{con1} \frac{H_{end}}{m_{pl}}>\beta_0^{5/2}g_*^{-1/2}\simeq
10^{-6}
\end{equation}
Taking the present radiation density fraction of the Universe to be
$\Omega_{\gamma_0}=2.5\cdot 10^{-5}h^{-2}$ ($h$ being the Hubble
constant in the units of $100km\cdot s^{-1}Mpc^{-1}$), and using the
standard values for the present time and time when the density of
matter and radiation become equal, we find the contemporary
densities fraction of relics
\begin{equation}
\label{reldens} \Omega_{rel}\approx 10^{26}h^{-2}
k\left(\frac{H_{end}}{m_{pl}}\right)^{3/2}
\end{equation}
It is easily to see that relics overclose the Universe
($\Omega_{rel}>>1$) for any reasonable $k$ and
$H_{end}>10^{-6}m_{pl}$.

The second case takes place if the Universe becomes PBHs dominated
at period $t_1<t_2<\tau_{BH}$. This situation is realized under the
condition  $\rho_{BH}(t_2)>\rho_{\gamma}(t_2)$, which can be
rewritten in the form
\begin{equation}
\label{con2} \frac{H_{end}}{m_{pl}}<10^{-6}.
\end{equation}
The present day relics density fraction takes the form
\begin{equation}
\label{reldens2} \Omega_{rel}\approx 10^{28}h^{-2}
k\left(\frac{H_{end}}{m_{pl}}\right)^{3/2}
\end{equation}
Thus the Universe is not overclosed by relics only if the following
condition is valid
\begin{equation}
\label{con3} \frac{H_{end}}{m_{pl}}\le 2\cdot
10^{-19}h^{4/3}k^{-2/3}.
\end{equation}
This condition implies that the masses of PBHs created at the end of
inflation have to be larger than
\begin{equation}
\label{massr} M_0\ge 10^{11}g\cdot h^{-4/3}\cdot k^{2/3}.
\end{equation}
From the other hand there are a number of well--known cosmological
and astrophysical limits \cite{15,mujana,151,152,153,154,155} which
prohibit the creation of PBHs in the mass range (\ref{massr}) with
initial fraction of mass density close to $\beta_0\approx 10^{-2}$.

So one have to conclude that the effect of the false vacuum bag
mechanism of PBH formation makes impossible the coexistence of
stable remnants of PBH evaporation with the first order phase
transitions at the end of inflation.

\section{Gravitino production by PBH evaporation and constraints
on the inhomogeneity of the early Universe}\label{gravitino}

Presently there are no observational evidences, proving existence of
PBHs. However, even the absence of PBHs provides a very sensitive
theoretical tool to study physics of early Universe. PBHs represent
nonrelativistic form of matter and their density decreases with
scale factor $a$ as $\propto a^{-3} \propto T^{3}$, while the total
density is $\propto a^{-4} \propto T^{4}$ in the period of radiation
dominance (RD). Being formed within horizon, PBH of mass $M$, can be
formed not earlier than at
\begin{equation}\label{tfRD}t(M)=\frac{M}{m_{pl}}{t_{pl}}=\frac{M}{m_{pl}^2}.\end{equation} If they are formed
on RD stage, the smaller are the masses of PBHs, the larger becomes
their relative contribution to the total density on the modern MD
stage. Therefore, even the modest constraint for PBHs of mass $M$ on
their density \beq
\label{OmPBH}\Omega_{PBH}(M)=\frac{\rho_{PBH}(M)}{\rho_{c}}\eeq in
units of critical density $\rho_{c}=3 H^2/(8 \pi G)$ from the
condition that their contribution $\alpha(M)$ into the the total
density
\begin{equation}\label{defalpha}\alpha(M)\equiv\frac{\rho_{PBH}(M)}{\rho_{tot}}=\Omega_{PBH}(M)
\end{equation} for $\rho_{tot}=\rho_{c}$ does not exceed the density
of dark
matter\begin{equation}\label{DMalpha}\alpha(M)=\Omega_{PBH}(M) \le
\Omega_{DM}=0.23\end{equation} converts into a severe constraint on
this contribution
\begin{equation}\label{defbeta}\beta \equiv
\frac{\rho_{PBH}(M,t_f)}{\rho_{tot}(t_f)}\end{equation} in the
period $t_f$ of their formation. If formed on RD stage at
$t_f=t(M)$, given by (\ref{tfRD}), which corresponds to the
temperature $T_f=m_{pl}\sqrt{m_{pl}/M}$, PBHs contribute into the
total density in the end of RD stage at $t_{eq}$, corresponding to
$T_{eq}\approx 1 eV$,  by factor
$a(t_{eq})/a(t_f)=T_f/T_{eq}=m_{pl}/T_{eq}\sqrt{m_{pl}/M}$ larger,
than in the period of their formation. The constraint on $\beta(M)$,
following from Eq.(\ref{DMalpha}) is then given
by\begin{equation}\label{DMbeta}\beta(M)=\alpha(M)\frac{T_{eq}}{m_{pl}}\sqrt{\frac{M}{m_{pl}}}
\le 0.23 \frac{T_{eq}}{m_{pl}}\sqrt{\frac{M}{m_{pl}}}.\end{equation}

The possibility of PBH evaporation, revealed by S. Hawking
\cite{hawking4}, strongly influences effects of PBHs. In the strong
gravitational field near gravitational radius $r_g$ of PBH quantum
effect of creation of particles with momentum $p \sim 1/r_g$ is
possible. Due to this effect PBH turns to be a black body source of
particles with temperature (in the units
$\hbar=c=k=1$)\begin{equation}\label{TPBHev}T=\frac{1}{8\pi G
M}\approx10^{13} {\rm GeV} \frac{1 {\rm g}}{M}.\end{equation} The
evaporation timescale BH is $\tau_{BH} \sim M^3/m_{pl}^4$ (see
Eq.(\ref{evop}) and discussion in previous section) and at $M \le
10^{14}$~g is less, than the age of the Universe. Such PBHs can not
survive to the present time and the magnitude Eq.(\ref{DMalpha}) for
them should be re-defined and has the meaning of contribution to the
total density in the moment of PBH evaporation. For PBHs formed on
RD stage and evaporated on RD stage at $t<t_{eq}$ the relationship
Eq.(\ref{DMbeta}) between $\beta(M)$ and $\alpha(M)$ is given by
\cite{NovikovPBH,polnarev}
\begin{equation}\label{DMbetaRD}\beta(M)=\alpha(M)\frac{m_{pl}}{M}.\end{equation}
The relationship between $\beta(M)$ and $\alpha(M)$ has more
complicated form, if PBHs are formed on early dust-like stages
\cite{polnarev1,polnarev,khlopov6,book}, or such stages take place
after PBH formation\cite{khlopov6,book}. Relative contribution of
PBHs to total density does not grow on dust-like stage and the
relationship between $\beta(M)$ and $\alpha(M)$ depends on details
of a considered model. Minimal model independent factor
$\alpha(M)/\beta(M)$ follows from the account for enhancement,
taking place only during RD stage between the first second of
expansion and the end of RD stage at $t_{eq}$, since radiation
dominance in this period is supported by observations of light
element abundance and spectrum of CMB
\cite{polnarev1,polnarev,khlopov6,book}.

Effects of PBH evaporation make astrophysical data much more
sensitive to existence of PBHs.
 Constraining the abundance of primordial
black holes can lead to invaluable information on cosmological
processes, particularly as they are probably the only viable probe
for the power spectrum on very small scales which remain far from
the Cosmological Microwave Background (CMB) and Large Scale
Structures (LSS) sensitivity ranges. To date, only PBHs with initial
masses between $\sim 10^9$~g and $\sim 10^{16}$~g have led to
stringent limits (see {\it e.g.}
\cite{carr1,carrMG,LGreen,polnarev}) from consideration of the
entropy per baryon, the deuterium destruction, the $^4$He
destruction and the cosmic-rays currently emitted by the Hawking
process \cite{hawking4}. The existence of light PBHs should lead to
important observable constraints, either through the direct effects
of the evaporated particles (for initial masses between $10^{14}$~g
and $10^{16}$~g) or through the indirect effects of their
interaction with matter and radiation in the early Universe (for PBH
masses between $10^{9}$~g and $10^{14}$~g). In these constraints,
the effects taken into account are those related with known
particles. However, since the evaporation products are created by
the gravitational field, any quantum with a mass lower than the
black hole temperature should be emitted, independently of the
strength of its interaction. This could provide a copious production
of superweakly interacting particles that cannot not be in
equilibrium with the hot plasma of the very early Universe. It makes
evaporating PBHs a unique source of all the species, which can exist
in the Universe.

Following \cite{book,book3,khlopov6,khlopov7} and
\cite{lemoine,green1} (but in a different framework and using more
stringent constraints),
 limits on the mass fraction of black holes
at the time of their formation ($\beta \equiv
\rho_{PBH}/\rho_{tot}$) were derived in \cite{KBgrain} using the
production of gravitinos during the evaporation process. Depending
on whether gravitinos are expected to be stable or metastable, the
limits are obtained using the requirement that they do not overclose
the Universe and that the formation of light nuclei by the
interactions of $^4$He nuclei with nonequilibrium flux of D,T,$^3$He
and $^4$He does not contradict the observations. This approach is
more constraining than the usual study of photo-dissociation induced
by photons-photinos pairs emitted by decaying gravitinos. It opened
a new window for the upper limits on $\beta$ below $10^9$~g. The
cosmological consequences of the limits, obtained in \cite{KBgrain},
are briefly reviewed in the framework of three different scenarios:
a blue power spectrum, a step in the power spectrum and first order
phase transitions.

\subsection{Limits on the PBH density}\label{evaporation}

Several constraints on the density of PBHs have been derived in
different mass ranges assuming the evaporation of only standard
model particles~: for $10^9~{\rm g}<M<10^{13}~{\rm g}$ the entropy
per baryon at nucleosynthesis  was used \cite{mujana} to obtain
$\beta < (10^9~{\rm g}/M)$, for $10^9~{\rm g}<M<10^{11}~{\rm g}$ the
production of $n\bar{n}$ pairs at nucleosynthesis was used
\cite{153} to obtain $\beta  < 3\times 10^{-17} (10^9~{\rm
g}/M)^{1/2}$ , for $10^{10}~{\rm g}<M<10^{11}~{\rm g}$ deuterium
destruction was used \cite{152} to obtain $\beta  < 3\times 10^{-22}
(M/10^{10}~{\rm g})^{1/2}$, for $10^{11}~{\rm g}<M<10^{13}~{\rm g}$
spallation of $^4$He was used \cite{vainer,khlopov6} to obtain
$\beta  < 3\times 10^{-21} (M/10^9~{\rm g})^{5/2}$, for $M\approx
5\times 10^{14}~{\rm g}$ the gamma-rays and cosmic-rays were used
\cite{155,barrau} to obtain $\beta < 10^{-28}$. Slightly more
stringent limits were obtained in \cite{kohri}, leading to $\beta <
10^{-20}$ for masses between $10^{9}~{\rm g}$ and $10^{10}~{\rm g}$
and in \cite{barraugamma}, leading to $\beta < 10^{-28}$ for
$M=5\times 10^{11}~{\rm g}$. Gamma-rays and antiprotons were also
recently re-analyzed in \cite{barraupbar} and
\cite{Custodio:2002jv}, improving a little the previous estimates.
Such constraints, related to phenomena occurring after the
nucleosynthesis, apply only for black holes with initial masses
above $\sim 10^9$~g. Below this value, the only limits are the very
weak entropy constraint (related with the photon-to-baryon ratio)
and the constraint, assuming stable remnants of black holes forming
at the end of the evaporation mechanism as described in the previous
Section.

\begin{figure}
    \begin{center}
        \includegraphics[scale=0.7]{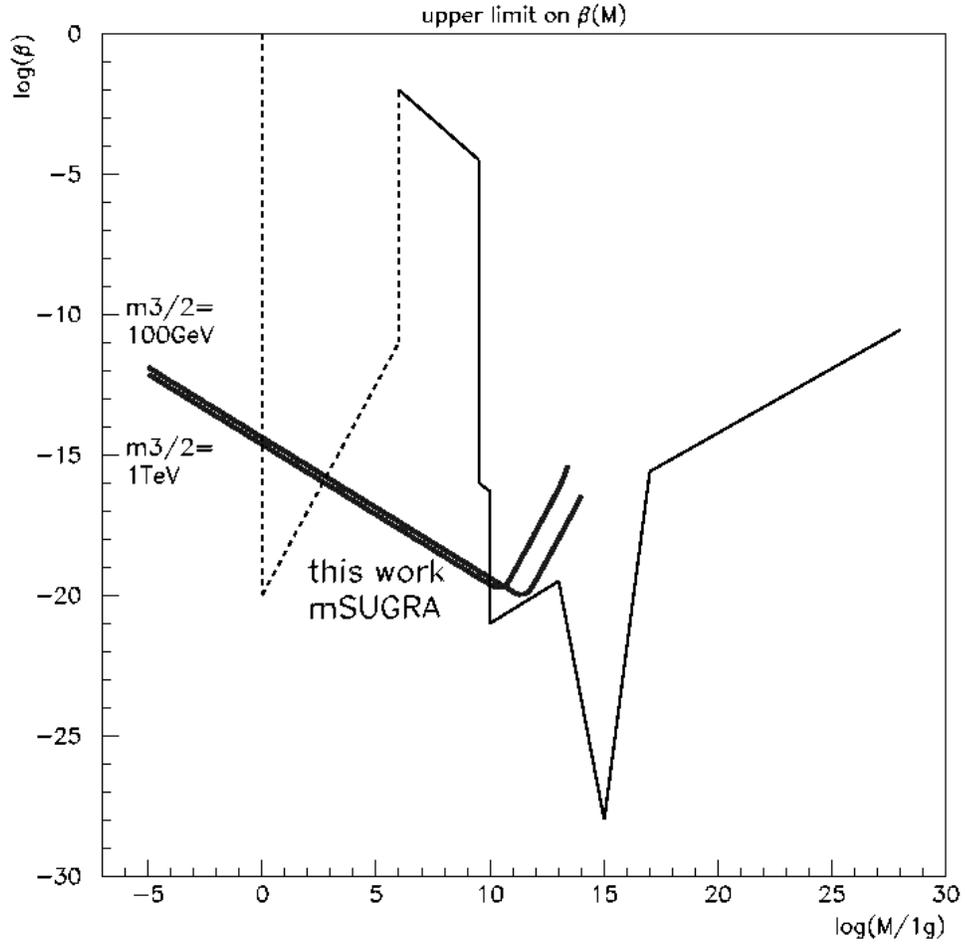}
        \caption{Constraints of \cite{KBgrain} on the fraction of the Universe going into PBHs (adapted from
\cite{carr1,carrMG,LGreen,polnarev}). The two curves obtained with
gravitinos emission in mSUGRA correspond to $m_{3/2}$ = 100 GeV
(lower curve in the high mass range) and  $m_{3/2}$ = 1 TeV (upper
curve in the high mass range)}
        \label{pot}
    \end{center}
\end{figure}

To derive a limit in the initial mass range $m_{pl}<M<10^{11}$~g,
gravitinos emitted by black holes were considered in \cite{KBgrain}.
Gravitinos are expected to be present in all local supersymmetric
models, which are regarded as the more natural extensions of the
standard model of high energy physics (see, {\it e.g.}, \cite{olive}
for an introductory review). In the framework of minimal
Supergravity (mSUGRA), the gravitino mass is, by construction,
expected to lie around the electroweak scale, {\it i.e.} in the 100
GeV range. In this case, the gravitino is {\it metastable} and
decays after nucleosynthesis, leading to important modifications of
the nucleosynthesis paradigm. Instead of using the usual
photon-photino decay channel, the study of \cite{KBgrain} relied on
the more sensitive gluon-gluino channel. Based on
\cite{khlopovlinde,khlopovlinde2,khlopovlinde3,khlopov3,khlopov31},
the antiprotons produced by the fragmentation of gluons emitted by
decaying gravitinos were considered as a source of nonequilibrium
light nuclei resulting from collisions of those antiprotons on
equilibrium nuclei. Then, $^6$Li, $^7$Li and $^7$Be nuclei
production by the interactions of the nonequilibrium nuclear flux
with $^4$He equilibrium nuclei was taken into account and compared
with data (this approach is supported by several recent analysis
\cite{Karsten,Kawasaki} which lead to similar results). The
resulting Monte-Carlo estimates \cite{khlopov3} lead to the
following constraint on the concentration of gravitinos: $n_{3/2}<
1.1\times 10^{-13}m_{3/2}^{-1/4}$, where $m_{3/2}$ is the gravitino
mass in GeV. This constraint has been successfully used to derive an
upper limit on the reheating temperature of the order
\cite{khlopov3}: $T_R < 3.8\times 10^6$~GeV. The consequences of
this limit on cosmic-rays emitted by PBHs was considered, {\it
e.g.}, in \cite{barrauprd}. In the approach of \cite{KBgrain} this
stringent constraint on the gravitino abundance was related to the
density of PBHs through the direct gravitino emission. The usual
Hawking formula \cite{hawking4} was used for the number of particles
of type $i$ emitted per unit of time $t$ and per unit of energy $Q$.
Introducing the  temperature defined by Eq. (\ref{TPBHev}) $
T=hc^3/(16\pi^2 k G M)\approx(10^{13}{\rm g})/{M}~{\rm GeV}, $
taking the relativistic approximation for $\Gamma_s$, and
integrating over time and energy, the total number of quanta of type
$i$ can be estimated as:
\begin{equation}\label{Niev}N_i^{TOT}=\frac{27\times 10^{24}}{64\pi ^3
\alpha_{SUGRA}}\int_{T_i}^{T_{Pl}}\frac{dT}{T^3}\int_{m/T}^x\frac{x^2dx}{e^x-(-1)^s}
\end{equation} where $T$ is in GeV, $m_{pl}\approx 10^{-5}$~g,
$x\equiv Q/T$, $m$ is the particle mass and $\alpha_{SUGRA}$
accounts for the number of degrees of freedom through
$M^2dM=-\alpha_{SUGRA}dt$ where $M$ is the black hole mass. Once the
PBH temperature is higher than the gravitino mass, gravitinos will
be emitted with a weight related with their number of degrees of
freedom. Computing the number of emitted gravitinos as a function of
the PBH initial mass and matching it with the limit on the gravitino
density imposed by nonequilibrium nucleosynthesis of light elements
leads to an upper limit on the PBH number density. If PBHs are
formed during a radiation dominated stage, this limit can easily be
converted into an upper limit on $\beta$ by evaluating the energy
density of the radiation at the formation epoch. The resulting limit
is shown on Fig. \ref{pot} and leads to an important improvement
over previous limits, nearly independently of the gravitino mass in
the interesting range. This opens a new window on the very small
scales in the early Universe.

\begin{figure}
    \begin{center}
        \includegraphics[scale=0.7]{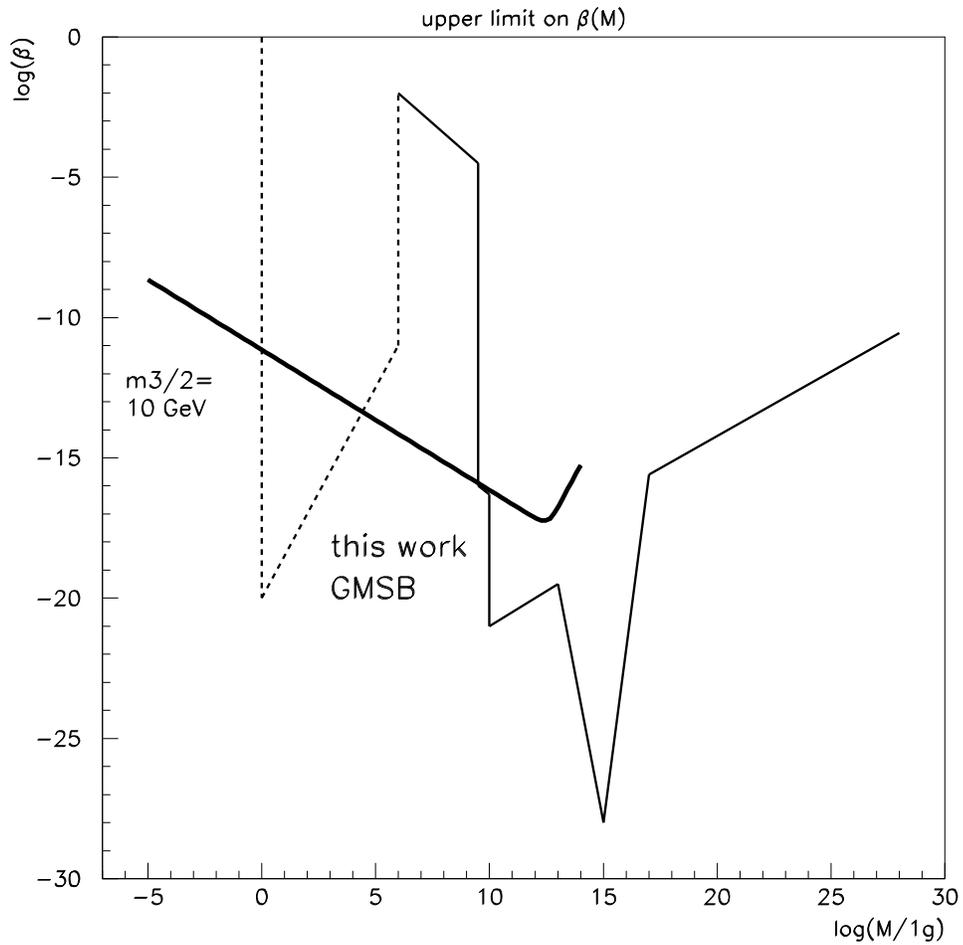}
        \caption{Constraints of \cite{KBgrain} on the fraction of the Universe going into PBHs (adapted from
\cite{carr1,carrMG,LGreen,polnarev}). The curve obtained with
gravitinos emission in GMSB correspond to $m_{3/2}=10$ GeV and
scales with gravitino mass as $\propto m_{3/2}^{-1}$.}
        \label{pot2}
    \end{center}
\end{figure}

It is also possible to consider limits arising in  Gauge Mediated
Susy Breaking (GMSB) models \cite{kolda}. Those alternative
scenarios, incorporating a natural suppression of the rate of
flavor-changing neutral-current due to the low energy scale, predict
the gravitino to be the Lightest Supersymmetric Particle (LSP). The
LSP is stable if R-parity is conserved. In this case, the limit was
obtained \cite{KBgrain} by requiring $\Omega_{3/2,0}<\Omega_{M,0}$,
{\it i.e.} by requiring that the current gravitino density  does not
exceed the matter density. It can easily be derived from the
previous method, by taking into account the dilution of gravitinos
in the period of PBH evaporation and conservation of gravitino to
specific entropy ratio, that \cite{KBgrain}:
\begin{equation}\label{BetDM}\beta \leq \frac{\Omega_{M,0}}{N_{3/2}\frac{m_{3/2}}{M}\left(
\frac{t_{eq}}{t_{f}} \right) ^{\frac{1}{2}}}\end{equation} where
$N_{3/2}$ is the total number of gravitinos emitted by a PBH with
initial mass $M$, $t_{eq}$ is the end of RD stage and
$t_f=\max(t_{form},t_{end})$ when a non-trivial equation of state for
the period of PBH formation is considered, {\it e.g.} a dust-like
phase which ends at $t_{end}$ \cite{polnarev1}. The limit
(\ref{BetDM}) does not imply thermal equilibrium of relativistic
plasma in the period before PBH evaporation and is valid even for
low reheating temperatures provided that the equation of state on
the preheating stage is close to relativistic. With the present
matter density $\Omega_{M,0}\approx 0.27$ \cite{wmap} this leads to
the limit shown in Fig. \ref{pot2} for $m_{3/2}=10$~GeV. Following
(\ref{BetDM}) this limit scales with gravitino mass as $\propto
m_{3/2}^{-1}$. Models of gravitino dark matter with $\Omega_{3/2,0}
= \Omega_{CDM,0}$, corresponding to the case of equality in the
above formula, were recently considered in
\cite{Jedamzik1,Jedamzik11}.

\subsection{Cosmological consequences}\label{spectrum}

Upper limits on the fraction of the Universe in primordial black
holes can be converted into cosmological constraints on models with
significant power on small scales \cite{KBgrain}.

The easiest way to illustrate the importance of such limits is to
consider a blue power spectrum and to derive a related upper value
on the spectral index $n$ of scalar fluctuations ($P(k)\propto
k^n$). It has recently been shown by WMAP \cite{wmap} that the
spectrum is nearly of the Harrison-Zel'dovich type, {\it i.e.} scale
invariant with $n\approx 1$. However this measure was obtained for
scales between $10^{45}$ and $10^{60}$ times larger that those
probed by PBHs and it remains very important to probe the power
available on small scales. The limit on $n$ given in \cite{KBgrain}
must therefore be understood as a way to constrain $P(k)$ at small
scales rather than a way to measure its derivative at large scales :
it is complementary to CMB measurements. Using the usual relations
between the mass variance at the PBH formation time
$\sigma_H(t_{form})$ and the same quantity today $\sigma_H(t_0)$
\cite{green},
\begin{equation}\label{Sigmh}\sigma_H(t_{form})=\sigma_H(t_0)\left(\frac{M_H(t_0)}{M_H(t_{eq})}\right)^{\frac{n-1}{6}}
\left(\frac{M_H(t_{eq})}{M_H(t_{form})}\right)^{\frac{n-1}{4}}\end{equation}
where $M_H(t)$ is the Hubble mass at time $t$ and $t_{eq}$ is the
equilibrium time, it is possible to set an upper value on $\beta$
which can be expressed as
\begin{equation}\label{betaspect}\beta\approx \frac{\sigma_H(t_{form})}{\sqrt{2\pi}\delta_{min}}e^{-\frac{\delta_{min}^2}
{2\sigma_H^2(t_{form})}},\end{equation} where $\delta_{min}\approx
0.3$ is the minimum density contrast required to form a PBH. The
limit derived in the previous subsection leads to $n<1.20$ in the
mSUGRA case whereas the usually derived limits range between 1.23
and 1.31 \cite{green,Bringmann:2001yp,kim2}. In the GMSB case, it
remains at the same level for $m_{3/2}\sim10$~GeV and is slightly
relaxed for smaller masses of gravitino. This improvement is due to
the much more important range of masses probed by the method
\cite{KBgrain}.

In the standard cosmological paradigm of inflation, the primordial
power spectrum is expected to be nearly --but not exactly-- scale
invariant \cite{liddle}. The sign of the running can, in principle,
be either positive or negative. It has been recently shown that
models with a positive running $\alpha_s$, defined as \beq
\label{specUV}P(k)=P(k_0)\left( \frac{k}{k_0}
\right)^{n_s(k_0)+\frac{1}{2}\alpha_s ln \left(
\frac{k}{k_0}\right)},\eeq are very promising in the framework of
supergravity inflation (see, {\it e.g.}, \cite{kawa}). The analysis
\cite{KBgrain} strongly limits a positive running, setting the upper
bound at a tiny value $\alpha_s<2\times 10^{-3}$. This result is
more stringent than the upper limit obtained through a combined
analysis of Ly$\alpha$ forest, SDSS and WMAP data \cite{seljak},
$-0.013<\alpha_s<0.007$, as it deals with scales very far from those
probed by usual cosmological observations. The order of magnitude of
the running naturally expected in most models --either inflationary
ones (see, {\it e.g.}, \cite{peiris}) or alternative ones (see, {\it
e.g.}, \cite{khoury})-- being of a few times $10^{-3}$ our upper
bound should help to distinguish between different scenarios.

In the case of an early dust-like stage in the cosmological
evolution \cite{khlopov0,polnarev,book,book3}, the PBH formation
probability is increased to $\beta > \delta ^ {13/2}$ where $\delta$
is the density contrast for the considered small scales (see
subsection \ref{particles}). The associated limit on $n$ is
strengthened to $n<1.19$.

\begin{figure}
    \begin{center}
        \includegraphics[scale=0.7]{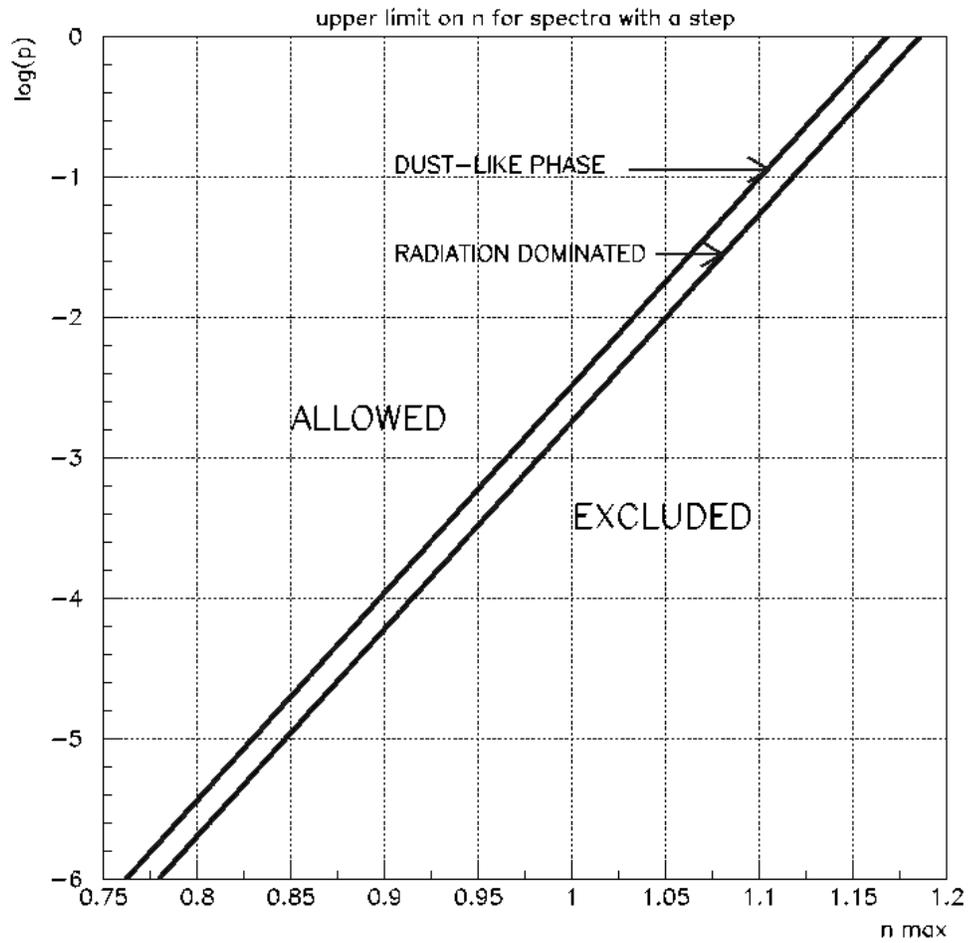}
        \caption{Upper limit from \cite{KBgrain} on the spectral index of the power spectrum as a function
of the amplitude of the step.}
        \label{pot3}
    \end{center}
\end{figure}

Following \cite{green}, it is also interesting to consider
primordial density perturbation spectra with both a tilt and a step.
Such a feature can arise from underlying physical processes
\cite{starobinsky} and allows investigation of a wider class of
inflaton potentials. If the amplitude of the step is defined so that
the power on small scales is $p^{-2}$ times higher than the power on
large scales, the maximum allowed value for the spectral index can
be computed as a function of $p$. Figure \ref{pot3}, taken from
\cite{KBgrain}, shows those limits, which become extremely stringent
when $p$ is small enough, for both the radiation-dominated and the
dust-like cases.

Another important consequence of limits \cite{KBgrain} concerns PBH
relics dark matter (see also discussion in subsection \ref{pt1}).
The idea, introduced in \cite{macgibbon2}, that relics possibly
formed at the end of the evaporation process could account for the
cold dark matter has been extensively studied. The amplitude of the
power boost required on small scales has been derived, {\it e.g.},
in \cite{barrau2} as a function of the relic mass and of the
expected density. The main point was that the "step" (or whatever
structure in the power spectrum) should occur at low masses to avoid
the constraints available between $10^9$~g and $10^{15}$~g. The
limit on $\beta$ derived in \cite{KBgrain} closes this dark matter
issue except within a small window below $10^3$~g.

This result can be re-formulated in a more general way. If the
nature of cosmological dark matter is related with superweakly
interacting particles, which can not be present in equilibrium in
early Universe and for which nonequilibrium processes of production
e.g. in reheating are suppressed, the early Universe should be
sufficiently homogeneous on small scales to exclude copious creation
of these species in miniPBH vaporation.

Finally, the limits \cite{KBgrain} also completely exclude the
possibility of a copious PBH formation process in bubble wall
collisions \cite{kkrs,kkrs1,kkrs2}, considered in the previous
Section. This has important consequences for the related constraints
on first order phase transitions in the early Universe and on
symmetry breaking pattern of particle theory.

\section {Massive Primordial Black Holes from collapse of closed walls}\label{MBHwalls}
A wide class of particle models possesses a symmetry breaking
pattern, which can be effectively described by
pseudo-Nambu--Goldstone (PNG) field and which corresponds to
formation of unstable topological defect structure in the early
Universe (see \cite{book2} for review and references). The
Nambu--Goldstone nature in such an effective description reflects
the spontaneous breaking of global U(1) symmetry, resulting in
continuous degeneracy of vacua. The explicit symmetry breaking at
smaller energy scale changes this continuous degeneracy by discrete
vacuum degeneracy. The character of formed structures is  different
for phase transitions, taking place on post-inflationary and
inflationary stages.
\subsection{Structures from succession of U(1) phase transitions}\label{structures}
At high temperatures such a symmetry breaking pattern implies the
succession of second order phase transitions. In the first
transition, continuous degeneracy of vacua leads, at scales
exceeding the correlation length, to the formation of topological
defects in the form of a string network; in the second phase
transition, continuous transitions in space between degenerated
vacua form surfaces: domain walls surrounded by strings. This last
structure is unstable, but, as was shown in the example of the
invisible axion \cite{Sakharov2,kss,kss2}, it is reflected in the
large scale inhomogeneity of distribution of energy density of
coherent PNG (axion) field oscillations. This energy density is
proportional to the initial value of phase, which acquires dynamical
meaning of amplitude of axion field, when axion mass is switched on
in the result of the second phase transition.

The value of phase changes by $2 \pi$ around string. This strong
nonhomogeneity of phase, leading to corresponding nonhomogeneity of
energy density of coherent PNG (axion) field oscillations, is usually
considered (see e.g. \cite{kim,Sikivie:2006ni} and references therein)
only on scales, corresponduing to mean distance between strings.
This distance is small, being of the order of the scale of
cosmological horizon in the period, when PNG field oscillations
start. However, since the nonhomogeneity of phase follows the
pattern of axion string network this argument misses large scale
correlations in the distribution of oscillations' energy density.

Indeed, numerical analysis of string network (see review in
\cite{vs}) indicates that large string loops are strongly suppressed
and the fraction of about 80\% of string length, corresponding to
long loops, remains virtually the same in all large scales. This
property is the other side of the well known scale invariant
character of string network. Therefore the correlations of energy
density should persist on large scales, as it was revealed in
\cite{Sakharov2,kss,kss2}.

The large scale correlations in topological defects and their
imprints in primordial inhomogeneities is the indirect effect of
inflation, if phase transitions take place after reheating of the
Universe. Inflation provides in this case equal conditions for phase
transition, taking place in causally disconnected regions.

If phase transitions take place on inflational stage new forms of
primordial large scale correlations appear. The value of phase after
the first phase transition is inflated over the region corresponding
to the period of inflation, while fluctuations of this phase
 change in
the course of inflation its initial value within the regions of
smaller size. Owing to such fluctuations, for the fixed value of
$\theta_{60}$ in the period of inflation with {\it e-folding} $N=60$
corresponding to the part of the Universe within the modern
cosmological horizon, strong deviations from this value appear at
smaller scales, corresponding to later periods of inflation with $N
< 60$. If $\theta_{60} < \pi$, the fluctuations can move the value
of $\theta_{N}$ to $\theta_{N} > \pi$ in some regions of the
Universe. After reheating in the result of the second phase
transition these regions correspond to vacuum with $\theta_{vac} =
2\pi$, being surrounded by the bulk of the volume with vacuum
$\theta_{vac} = 0$. As a result massive walls are
 formed at the border between the
two vacua. Since regions with $\theta_{vac} = 2\pi$ are confined,
the domain walls are closed. After their size equals the horizon,
closed walls can collapse into black holes.

  This mechanism can lead to formation
of primordial black holes of a whatever large mass (up to the mass
of AGNs \cite{AGN}, see for latest review \cite{DER}). Such black
holes appear in the form of primordial black hole clusters,
exhibiting fractal distribution in space
\cite{KRS,Khlopov:2004sc,book2}. It can shed new light on the
problem of galaxy formation \cite{book2,DER1,DER2}.

\subsection {Formation of closed walls in inflationary Universe}\label{walls}
To describe a mechanism for the appearance of massive walls of a
size essentially greater than the horizon at the end of inflation,
let us consider a complex scalar field with the
potential\cite{AGN,KRS,Khlopov:2004sc,book2}
\begin{equation}\label{V1} V(\varphi ) = \lambda (\left| \varphi
\right|^2  - f^2 /2)^2+\delta V(\theta ), \end{equation} where
$\varphi  = re^{i\theta } $. This field coexists with an inflaton
field which drives the Hubble constant $H$ during the inflational
stage. The term
\begin{equation} \label{L1} \delta V(\theta ) = \Lambda ^4 \left(
{1 - \cos \theta } \right), \end{equation} reflecting the
contribution of instanton effects to the Lagrangian renormalization
(see for example \cite{adams}), is negligible on the inflational
stage and during some period in the FRW expansion. The omitted term
(\ref{L1}) becomes significant, when temperature falls down the
values $T \sim \Lambda$. The mass of radial field component $r$ is
assumed to be sufficiently large with respect to $H$, which means
that the complex field is in the ground state even before the end of
inflation. Since the term (\ref{L1}) is negligible during inflation,
the field has the form $\varphi \approx f/\sqrt 2 \cdot e^{i\theta }
$, the quantity $f\theta$ acquiring the meaning of a massless field.


At the same time,  the well established behavior of quantum field
fluctuations on the de Sitter background \cite{Star80} implies that
the wavelength of a vacuum fluctuation of every scalar field grows
exponentially, having a fixed amplitude. Namely, when the wavelength
of a particular fluctuation, in the inflating Universe, becomes
greater than $H^{-1}$, the average amplitude of this fluctuation
freezes out at some  non-zero value because of the large friction
term in the equation of motion  of the scalar field, whereas its
wavelength grows exponentially. Such a frozen fluctuation is
equivalent to the appearance of a classical field that does not
vanish after averaging over macroscopic space intervals. Because the
vacuum must contain fluctuations of every wavelength, inflation
leads to the  creation of more and more new regions containing a
classical field of different amplitudes with scale greater than
$H^{-1}$. In the case of an effectively massless Nambu--Goldstone
field considered here, the averaged amplitude of phase fluctuations
generated during each e-fold (time interval $H^{-1}$)  is given by
\beq \label{fluctphase} \delta \theta = H/2\pi f. \eeq Let us assume
that the part of the Universe observed inside the contemporary
horizon $H_0^{-1}=3000h^{-1}$Mpc was inflating, over $N_U \simeq 60$
e-folds, out of a single causally connected domain of size $H^{-1}$,
which contains some average value of phase $\theta_0$ over it. When
inflation begins in this region, after one e-fold, the volume of the
Universe increases by a factor $e^3$ . The typical wavelength of the
fluctuation $\delta\theta$ generated during every e-fold is equal to
$H^{-1}$. Thus, the whole domain  $H^{-1}$, containing $\theta_{0}$,
after the first e-fold effectively becomes divided into  $e^3$
separate, causally disconnected domains of size $H^{-1}$. Each
domain contains almost homogeneous  phase value
$\theta_{0}\pm\delta\theta$. Thereby, more and more domains appear
with time, in which the phase differs significantly from the initial
value $\theta_0$. A principally important point is the appearance of
domains with phase $\theta >\pi$. Appearing only after a certain
period of time during which the Universe exhibited exponential
expansion, these domains turn out to be surrounded by a space with
phase $\theta <\pi$. The coexistence of domains with phases $\theta
<\pi$ and $\theta
>\pi$ leads, in the following, to formation of
a large-scale structure of topological defects.

The potential (\ref{V1}) possesses a $U(1)$ symmetry, which is
spontaneously broken, at least, after some period of inflation. Note
that the phase fluctuations during the first e-folds may, generally
speaking, transform eventually into fluctuations of the cosmic
microwave radiation, which will lead to imposing restrictions on the
scaling parameter $f$. This difficulty can be avoided by taking into
account the interaction of the field $\varphi$ with the inflaton
field (i.e. by making parameter $f$ a variable~\cite{book2}). This
spontaneous breakdown is holding by the condition on
the radial mass, $m_r=\sqrt{\lambda}f>H$. At the same time the
condition \beq\label{angularmass} m_{\theta}=\frac{2f}{\Lambda}^2\ll
H \eeq on the angular mass provides the freezing out  of the phase
distribution until some moment of the FRW epoch.  After the
violation of condition (\ref{angularmass}) the term (\ref{L1})
contributes significantly to the potential (\ref{V1}) and explicitly
breaks the continuous symmetry along the angular direction. Thus,
potential (\ref{V1}) eventually has a number of discrete degenerate
minima in the angular direction at the points $\theta_{min}=0,\ \pm
2\pi ,\ \pm 4\pi,\ ...$ .

As soon as the angular mass $m_{\theta}$ is of the order of the
Hubble rate, the phase starts oscillating about the potential
minimum, initial values being different in various space domains.
Moreover, in the domains with the initial phase $\pi <\theta < 2\pi
$, the oscillations proceed around the potential minimum at $\theta
_{min}=2\pi$, whereas the phase in the surrounding space tends to a
minimum at the point $\theta _{min}=0$. Upon ceasing of the decaying
phase oscillations, the system contains domains characterized by the
phase $\theta _{min}=2\pi$ surrounded by space with $\theta
_{min}=0$. Apparently, on moving in any direction from inside to
outside of the domain, we will unavoidably pass through a point
where $\theta =\pi$ because the phase varies continuously. This
implies that a closed surface characterized by the phase $\theta
_{wall}=\pi$ must exist. The size of this surface depends on the
moment of domain formation in the inflation period, while the shape
of the surface may be arbitrary. The principal point for the
subsequent considerations is that the surface is closed. After
reheating of the Universe, the evolution of domains with the phase
$\theta >\pi $ proceeds on the background of the Friedman expansion
and is described by the relativistic equation of state. When the
temperature falls down to $T^* \sim \Lambda$, an equilibrium state
between the "vacuum" phase $\theta_{vac}=2\pi$ inside the domain and
the $\theta_{vac} =0$ phase outside it is established. Since the
equation of motion corresponding to potential (\ref{L1}) admits a
kink-like solution (see \cite{vs} and references therein), which
interpolates between two adjacent vacua $\theta_{vac} =0$  and
$\theta_{vac} =2\pi$,  a closed wall corresponding to the transition
region at $\theta =\pi$ is formed. The surface energy density of a
wall of width $\sim 1/m\sim  f/\Lambda^2$ is of the order of $\sim
f \Lambda ^2$ \footnote{The existence of such domain walls in theory
of the invisible axion was first pointed out in
\cite{sikivieinvisible}.}.

Note that if the coherent phase oscillations do not decay for a long
time, their energy density can play the role of CDM. This is the
case, for example, in the cosmology of the invisible axion (see
\cite{kim,Sikivie:2006ni} and references therein).

It is clear that immediately after the end of inflation, the size of
domains which contains a phase $\theta_{vac} >2\pi$ essentially
exceeds the horizon size.  This situation is replicated in the size
distribution of vacuum walls, which appear at the temperature $T^*
\sim \Lambda$ whence the angular mass $m_{\theta}$ starts to build
up. Those walls, which are larger than the cosmological horizon,
still follow the general FRW expansion until the moment when they
get causally connected as a whole; this happens as soon as the size
of a wall becomes equal to the horizon size $R_h$. Evidently,
internal stresses developed in the wall after crossing  the horizon
initiate processes tending to minimize the wall  surface. This
implies that the wall tends, first, to acquire a  spherical shape
and, second, to contract toward the centre. For simplicity, we will
consider below the motion of closed spherical walls~\footnote{The
motion of closed vacuum walls has been driven analytically in
\cite{tkachev,sikivie}.}.

The wall energy is proportional to its area at the instant of
crossing the horizon. At the moment of maximum contraction, this
energy is almost completely converted into kinetic energy
\cite{Rubinwall}. Should the wall at the same moment be localized
within the gravitational radius, a PBH is formed.

Detailed consideration of BH formation was performed in \cite{AGN}.
The results of these calculations are sensitive to changes in the
parameter $\Lambda$ and the initial phase $\theta _U$. As the
$\Lambda$ value decreases to $\approx 1$GeV, still greater PBHs
appear with masses of up to $\sim 10^{40}$ g. A change in the
initial phase leads to sharp variations in the total number of black
holes.As was shown above, each domain generates a family of
subdomains in the close vicinity. The total mass of such a cluster
is only 1.5--2 times that of the largest initial black hole in this
space region. Thus, the calculations confirm the possibility of
formation of clusters of massive PBHs ( $\sim 100M_{\odot}$ and
above) in the pregalactic stages of the evolution of the Universe. 
These clusters represent stable
energy density fluctuations around which increased baryonic (and
cold dark matter) density may concentrate in the subsequent stages,
followed by the evolution into galaxies.

It should be noted that additional energy density is supplied by
closed walls of small sizes. Indeed, because the smallness of their
gravitational radius, they do not collapse into BHs. After several
oscillations such walls disappear, leaving coherent fluctuations of
the PNG field. These fluctuations contribute to a local energy
density excess, thus facilitating the formation of galaxies.

The mass range of formed BHs is constrained by fundamental
parameters of the model $f$ and $\Lambda$. The maximal BH mass is
determined by the condition that the wall does not dominate locally
before it enters the cosmological horizon. Otherwise, local wall
dominance leads to a superluminal $a \propto t^2$ expansion for the
corresponding region, separating it from the other part of the
Universe. This condition corresponds to the mass \cite{book2}\beq
\label{Mmax} M_{max} =
\frac{m_{pl}}{f}m_{pl}(\frac{m_{pl}}{\Lambda})^2.\eeq The minimal
mass follows from the condition that the gravitational radius of BH
exceeds the width of wall and it is equal to\cite{KRS,book2}\beq
\label{Mmin} M_{min} = f(\frac{m_{pl}}{\Lambda})^2.\eeq

Closed wall collapse leads to primordial GW spectrum, peaked at \beq
\label{nupeak}\nu_0=3\cdot 10^{11}(\Lambda/f){\rm Hz} \eeq with
energy density up to \beq \label{OmGW}\Omega_{GW} \approx
10^{-4}(f/m_{pl}).\eeq At $f \sim 10^{14}$GeV this primordial
gravitational wave background can reach $\Omega_{GW}\approx
10^{-9}.$ For the physically reasonable values of \beq
1<\Lambda<10^8{\rm GeV}\eeq the maximum of spectrum corresponds to
\beq 3\cdot 10^{-3}<\nu_0<3\cdot 10^{5}{\rm Hz}.\eeq Another
profound signature of the considered scenario are gravitational wave
signals from merging of BHs in PBH cluster. These effects can
provide test of the considered approach in LISA experiment.
\section{Discussion}\label{Discussion}
For long time scenarios with Primordial Black holes belonged
dominantly to cosmological {\it anti-Utopias}, to "fantasies", which
provided restrictions on physics of very early Universe from
contradiction of their predictions with observational data. Even
this "negative" type of information makes PBHs an important
theoretical tool. Being formed in the very early Universe as
initially nonrelativistic form of matter, PBHs should have increased
their contribution to the total density during RD stage of
expansion, while effect of PBH evaporation should have strongly
increased the sensitivity of astrophysical data to their presence.
It links astrophysical constraints on hypothetical sources of cosmic
rays or gamma background, on hypothetical factors, causing influence
on light element abundance and spectrum of CMB, to restrictions on
superheavy particles in early Universe and on first and second order
phase transitions, thus making a sensitive astrophysical probe to
particle symmetry structure and pattern of its breaking at superhigh
energy scales.

Gravitational mechanism of particle creation in PBH evaporation
makes evaporating PBH an unique source of any species of particles,
which can exist in our space-time. At least theoretically, PBHs can
be treated as source of such particles, which are strongly
suppressed in any other astrophysical mechanism of particle
production, either due to a very large mass of these species, or
owing to their superweak interaction with ordinary matter.

By construction astrophysical constraint excludes effect, predicted
to be larger, than observed. At the edge such constraint converts
into an alternative mechanism for the observed phenomenon. At some
fixed values of parameters, PBH spectrum can play a positive role
and shed new light on the old astrophysical problems.

The common sense is to think that PBHs should have small sub-stellar
mass. Formation of PBHs within cosmological horizon, which was very
small in very early Universe, seem to argue for this viewpoint.
However, phase transitions on inflationary stage can provide spikes
in spectrum of fluctuations at any scale, or provide formation of
closed massive domain walls of any size.

In the latter case primordial clouds of massive black holes around
intermediate mass or supermassive black hole is possible. Such
clouds have a fractal spatial distribution. A development of this
approach gives ground for a principally new scenario of the galaxy
formation in the model of the Big Bang Universe. Traditionally, Big
Bang model assumes a homogeneous distribution of matter on all
scales, whereas the appearance of observed inhomogeneities is
related to the growth of small initial density perturbations.
However, the analysis of the cosmological consequences of the
particle theory indicates the possible existence of strongly
inhomogeneous primordial structures in the distribution of both the
dark matter and baryons. These primordial structures represent a new
factor in galaxy formation theory. Topological defects such as the
cosmological walls and filaments, primordial black holes, archioles
in the models of axionic CDM, and essentially inhomogeneous
baryosynthesis (leading to the formation of antimatter domains in
the baryon-asymmetric Universe
\cite{exl1,exl2,crg,kolb,we,khl,CSKZ,zil,sb,dolgmain,
khlopgolubkov,bgk,FarKhl,book,book2,book3}) offer by no means a
complete list of possible primary inhomogeneities inferred from the
existing elementary particle models.

Observational cosmology offers strong evidences favoring the
existence of processes, determined by new physics, and the
experimental physics approaches to their investigation.
Cosmoparticle physics \cite{ADS,MKH,book,book3}, studying the
physical, astrophysical and cosmological impact of new laws of
Nature, explores the new forms of matter and their physical
properties. Its development offers the great challenge for
theoretical and experimental research. Physics of Primordial Black
holes can play important role in this process.
\section*{Acknowledgement}
I express my gratitude to J.A. de Freitas Pacheco for inviting me to
write this review and I am grateful to A. Barrau, S.G.Rubin and
A.S.Sakharov for discussions and help in preparation of manuscript.

\end{document}